# VHDL IMPLEMENTATION AND VERIFICATION OF ARINC-429 CORE

***Prof.M.Kamaraju, * Dr A.V.N.Tilak, **Dr K.Lalkishore, *K.Baburao**

*Gudlavalleru EngineeringCollege, Gudlavalleru, **JNTU, Hyderabad

madduraju@yahoo.com

**ABSTRACT**

Modern Avionics are controlled by sophisticated mission components in the Aircraft. The control function is implemented via a standard ARINC-429 bus interface. It is a two-wire point-to-point serial data bus for control communications in Avionics. The bus operates 12.5 or 100kb/sec, the implementation is envisaged for one transmits and receive channel respectively. Further the code can be modified for more no of independent Tx and Rx channels. An on chip memory allotment on the FPGA will provide a buffer bank for storing the incoming or outgoing data. For this purpose SRAM based FPGAs are utilized. This flexible ARINC429 solution gives exactly what is needed for real time applications. The IP can be programmed to send an interrupt to the host and also prepare it to process the data. Majority of the hardware function of digital natures are embedded into a single FPGA by saving in terms of PCB board space, power consumption and volume results. This paper deals with the development, implementation, simulation, and verification of ARINC_429 formats. The IP core development is described in VHDL.

# 1.INTRODUCTION

### 1.1 ARINC 429 Overview

ARINC 429 is a two-wire; point-to-point data bus that is application-specific for commercial and transport aircraft. The connection wires are twisted pairs. Words are 32 bits in length and most messages consist of a single data word. The specification defines the electrical standard and data characteristics and protocols. ARINC 429 uses a unidirectional data bus standard (Tx and Rx are on separate ports) known as the Mark 33 digital Information Transfer System (DITS). Messages are transmitted at 12.5, 50 (optional), or 100 kbps to other system elements that are monitoring the bus messages. The transmitter is always transmitting either 32-bit data words or the Null state. Each ARINC word contains five fields: Parity, Sign/Status Matrix, Data, and Source/Destination Identifiers, Label. In Fig.1, the parity bit is bit 32 (the MSB), SSM is the Sign/Status Matrix and is included as bits 30 and 31, bits 11 to 29 contain the data. Binary Coded Decimal (BCD) and binary encoding (BNR) are common ARINC data formats. Data formats can also be mixed. Bits 9 and 10 are Source/Destination Identifiers (SDI) and indicate for which receiver the data is intended. Bits 1 to 8 contain a label (label words) identifying the data type. Label words are quite specific in ARINC 429. Each aircraft may be equipped with different electronic equipment and systems needing interconnection. A large amount of equipment may be involved, depending on the aircraft. The ARINC specification identifies the equipment ID, a series of digital identification numbers. Examples of equipment are Flight Management Computers, Inertial Reference Systems, Fuel Tanks, Tire Pressure Monitoring Systems, and GPS Sensors.

| 32 | 31 | 30 | 29   …   11 | 10 | 9 | 8   …   1 |
|---|---|---|---|---|---|---|
| P | SSM | | DATA → ← PAD ← DISCRETES<br>MSB   LSB | SDI | | LABEL |

Fig. 1 *ARINC* Data Bit Positions

### 1.2 Transmission Order

The least significant bit of each byte, except the label, is transmitted first, and the label is transmitted ahead of the Data in each case. The order of the bits transmitted on the ARINC bus is as follows: 8, 7, 6, 5, 4, 3, 2, 1, 9, 10, 11, 12, 13 … 32.

# 2. Core 429 Overview

Core 429 provides a complete and flexible interface to a microprocessor and an ARINC 429 data bus. Connection to an ARINC 429 data bus requires additional line drivers and line receivers.Core429 interfaces to a processor through the internal memory of the receiver. Core429 can be easily interfaced to an 8-, 16- or 32-bit data bus. Look-up tables loaded into memory enable the Core429 receive circuitry to filter and sort incoming data by label and destination bits. Core429 supports multiple (configurable) ARINC 429 receiver channels, and each receives data independently. The receiver data rates (high or low speed) can be programmed independently. Core429 can decode and sort data based on the ARINC 429 Label and SDI bits and stores it in FIFO. Each receiver uses programmable FIFO to buffer received data. Core429 supports multiple (configurable) ARINC 429 transmit channels and each channel can transmit data independently.

### 2.1 Functional Description

The core has three main blocks: Transmitter, Receiver, and CPU interface. The core can be configured to provide up to 16 transmit and receive channels.

**2.1 (a) Rx Block**: The Fig.2, Rx block is responsible for recovering the clock from the input serial data and performs serial-to-parallel Conversion and gap/parity check on the incoming data. It also interfaces with the CPU. The Rx module contains two 8-bit registers. One is used for control function and the other is used for status. The CPU interface configures the internal RAM with the labels, which are used to compare against the incoming labels from the received ARINC data.If the label-compare bit in the receive control register is enabled, then the data which matches its labels with the stored labels will be stored in the FIFO. If the label compare bit in the receive control register is disabled, then the incoming data will be stored in the FIFO without comparing against the labels in RAM.

The core supports reloading label memory using bit 7 of the Rx control register,set bit 7 to initialize the label memory, the old label content still exists, but the core keeps

.

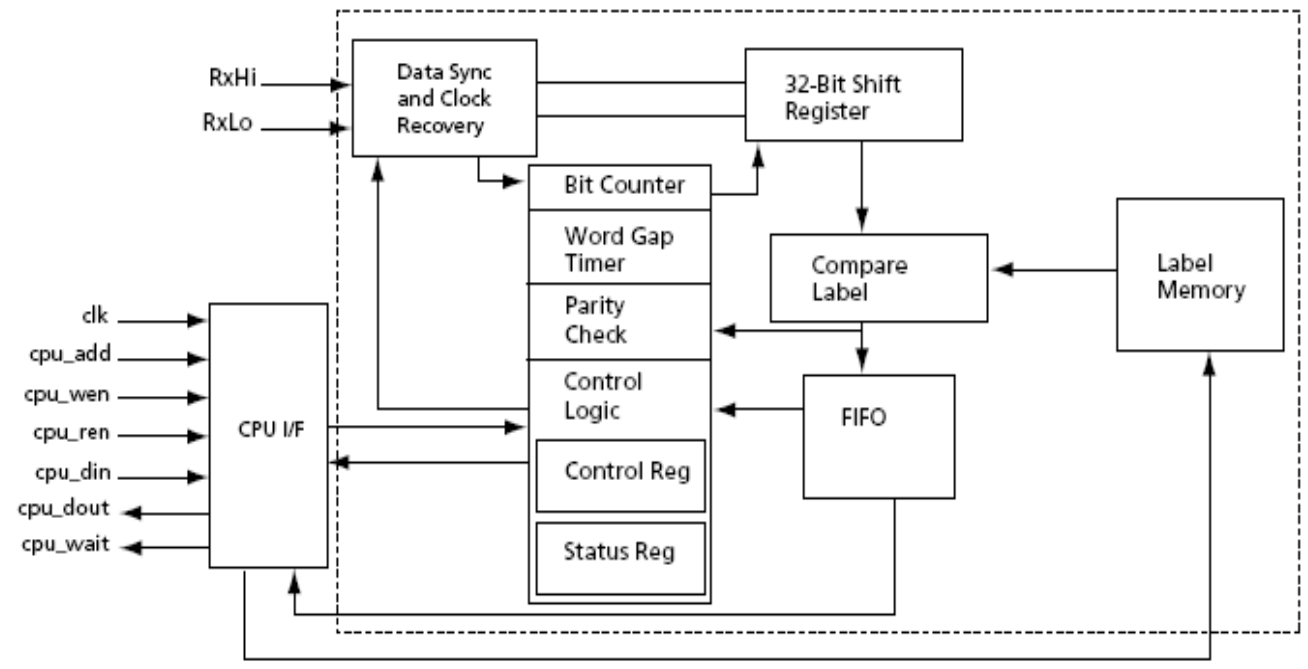

Fig.2 *Core429* Rx (receiver) block Diagram

track only of the new label and does not use the old label during label compare. The FIFO asserts three status signals: rx_fifo_empty: FIFO is empty, rx_fifo_half_full: FIFO is filled up to the programmed RX_FIFO_LEVEL, rx_fifo_full: FIFO is full Depending on the FIFO status signals, the CPU will either read the FIFO before it overflows, or not attempt to read the FIFO if it is empty. The interrupt signal int_out_rx is generated when one of the FIFO status signals (rx_fifo_empty, rx_fifo_half_full, and rx_fifo_full) are high.

**2.1(b) Tx Block**: Fig. 3 shows core 429 Tx bock converts the 32-bit parallel data from the Tx FIFO to serial data. It also inserts the parity bit into the ARINC data when parity is enabled. The CPU interface is used to fill the FIFO with ARINC data. The Tx FIFO can hold up to 512 ARINC words of data. The transmission starts as soon as one complete ARINC word has been stored in the transmit FIFO.

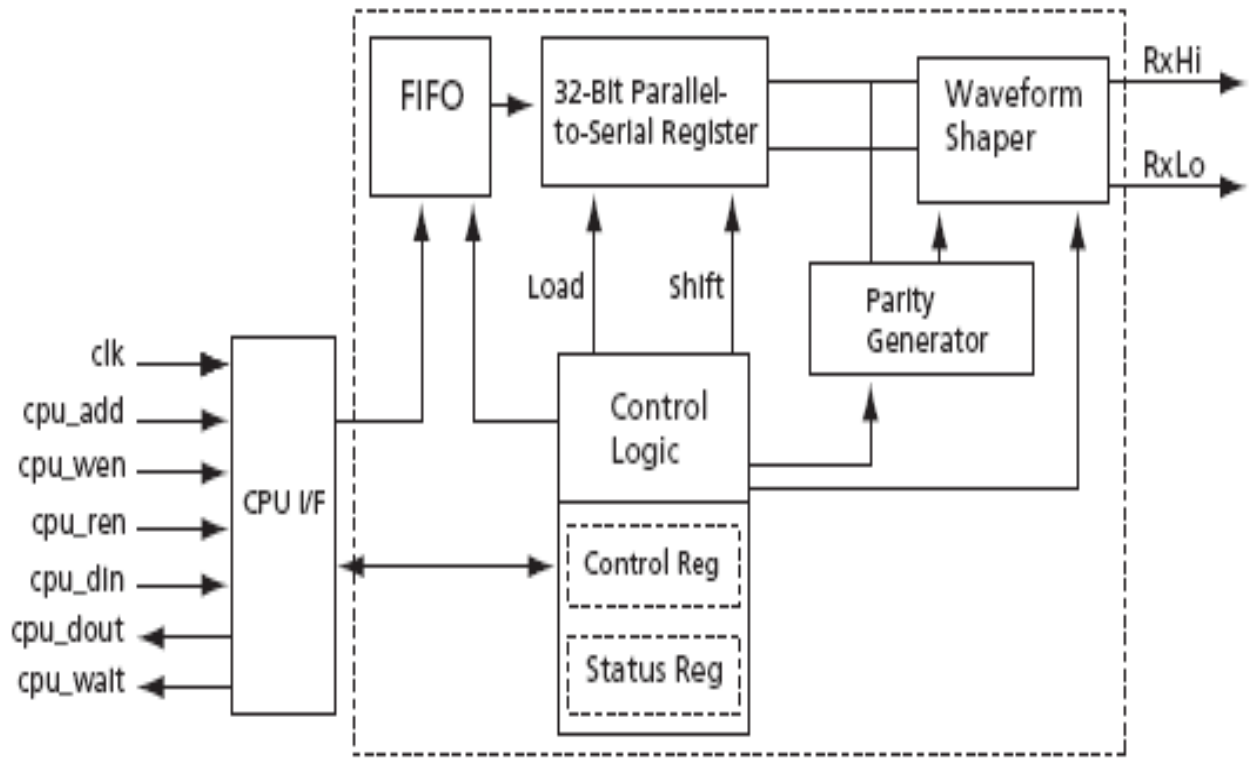

Fig.3 Core 429 Tx (transmitter) block diagram

The Tx module contains two 8-bit registers. One is used for a control function and the other is used for status. The CPU interface allows the system CPU to access the control and status registers within the core. The Tx FIFO asserts three status signals: tx_fifo_empty: Tx FIFO is empty, tx_fifo_half_full: Tx FIFO is filled up to the programmed TX_FIFO_LEVEL, tx_fifo_full: TX FIFO is full Depending on the FIFO status signals, the CPU will either read the FIFO before it overflows, or not attempt to read the FIFO if it is empty. The interrupt signal int_out_tx is generated when one of the FIFO status signals (tx_fifo_empty, tx_fifo_half_full and tx_fifo_full) are high.

**2.1 (c) CPU Interface**

The CPU interface allows access to the Core429 internal registers, FIFO, and internal memory. This interface is synchronous to the clock.

| Name | Type | Description |
|---|---|---|
| cpu_ren | In | CPU read enable, active low |
| cpu_wen | In | CPU write enable, active low |
| cpu_add [8:0] | In | CPU address |
| cpu_din [CPU_DATA_WIDTH-1:0] | In | CPU data input |
| cpu_dout [CPU_DATA_WIDTH-1:0] | Out | CPU data output |
| int_out | Out | Interrupt to CPU, active high. int_out is the OR function of int_out_rx and int_out_tx. |
| cpu_wait | Out | Indicates that the CPU should hold cpu_ren or cpu_wen active while the core completes the read or write operation. |

.

Table 1 CPU Interface Signals

## 3. Implementation

### 3.1 VLSI Design Flow

The whole design is implemented in VHDL using Altera Quartus II software. The following steps are followed in the whole design. Make use of Cyclone targated board (Fig.5),ARNIC core 429 tested and verified successfully.

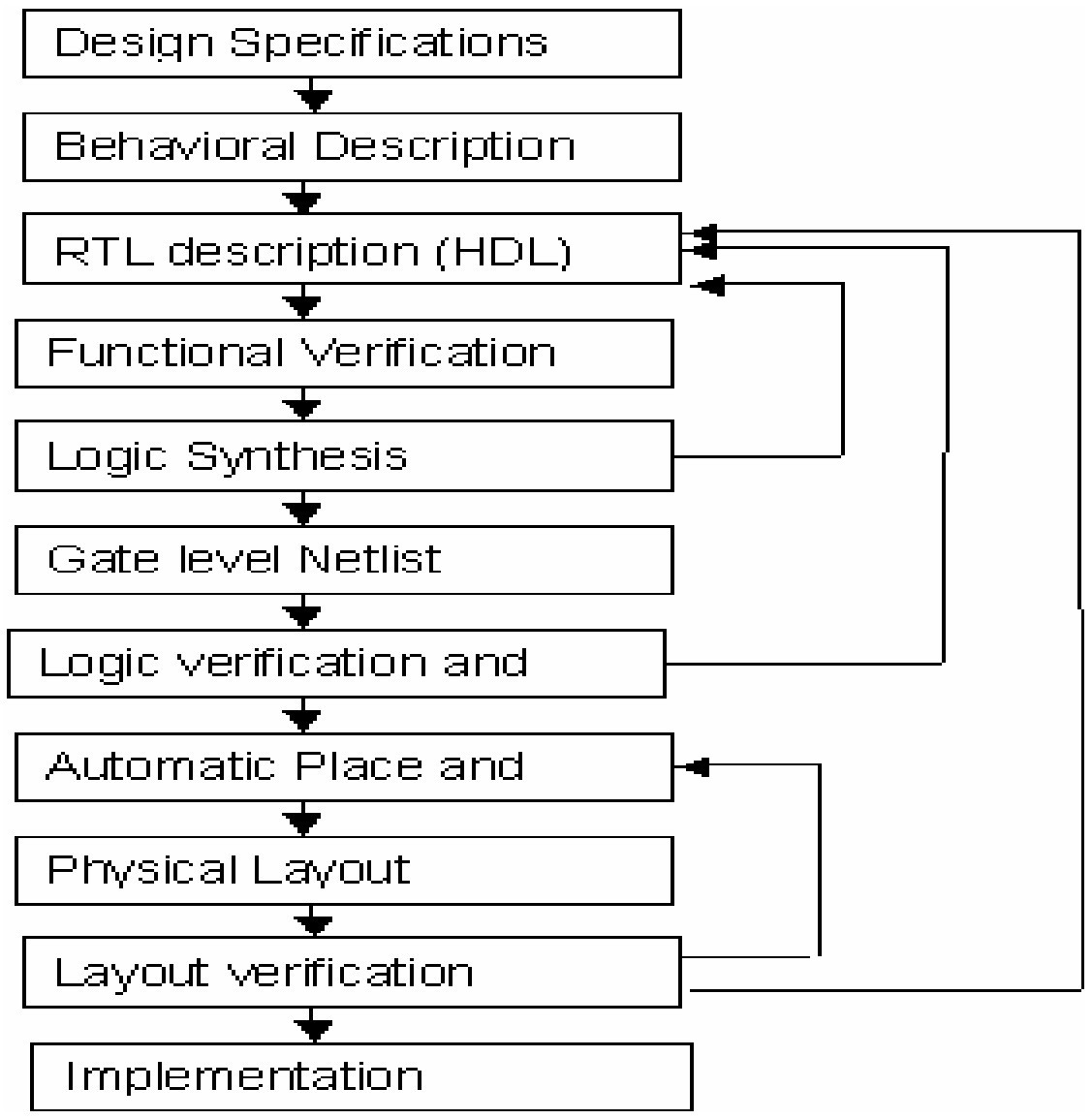

F 4 VLSI Design Flow

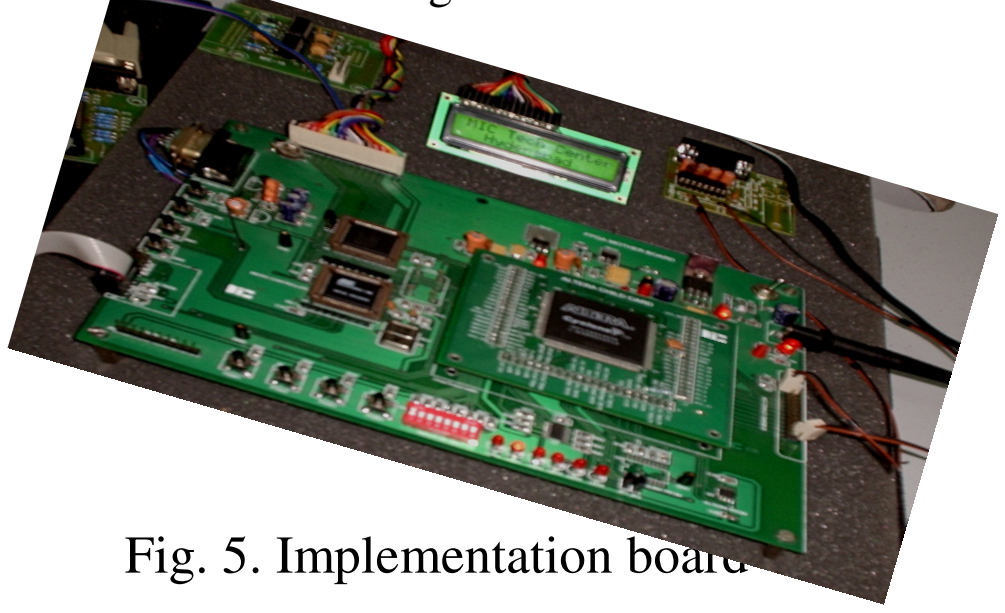

Fig. 5. Implementation board

# 5.Results

## 5.1 SIMULATION REPORTS

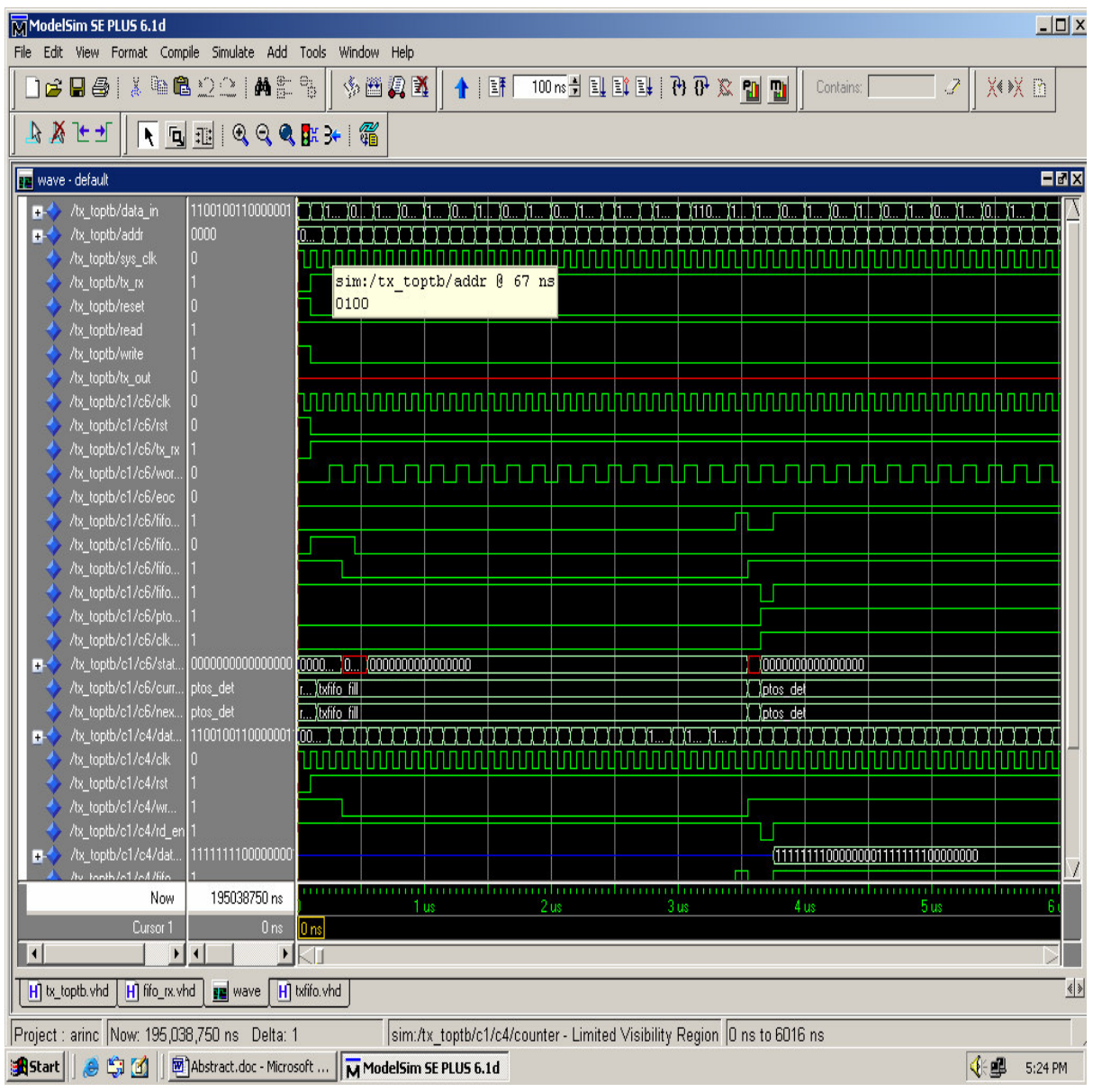

Fig.7a. Transmitter Top Level Results 1

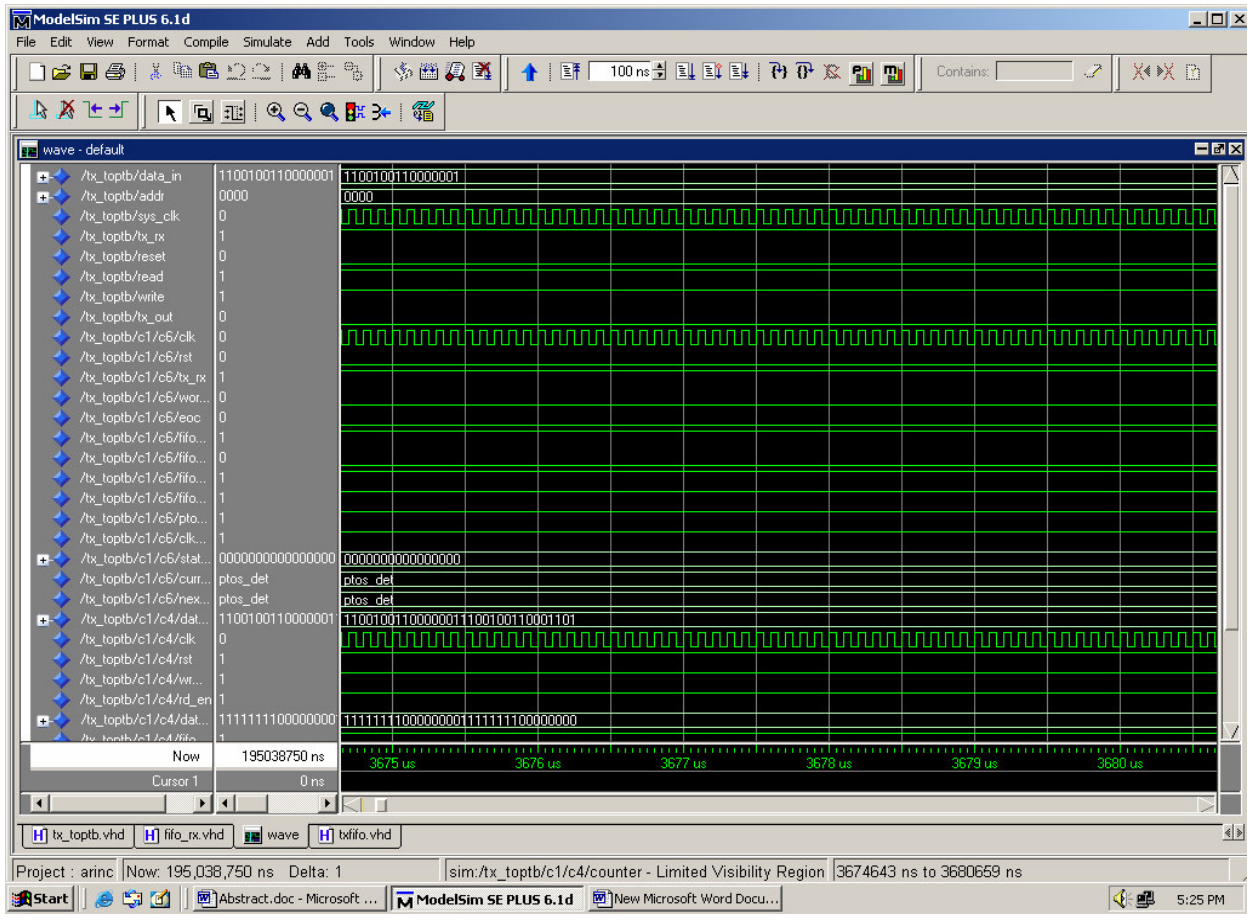

Fig.7b. Transmitter Top Level Results 2

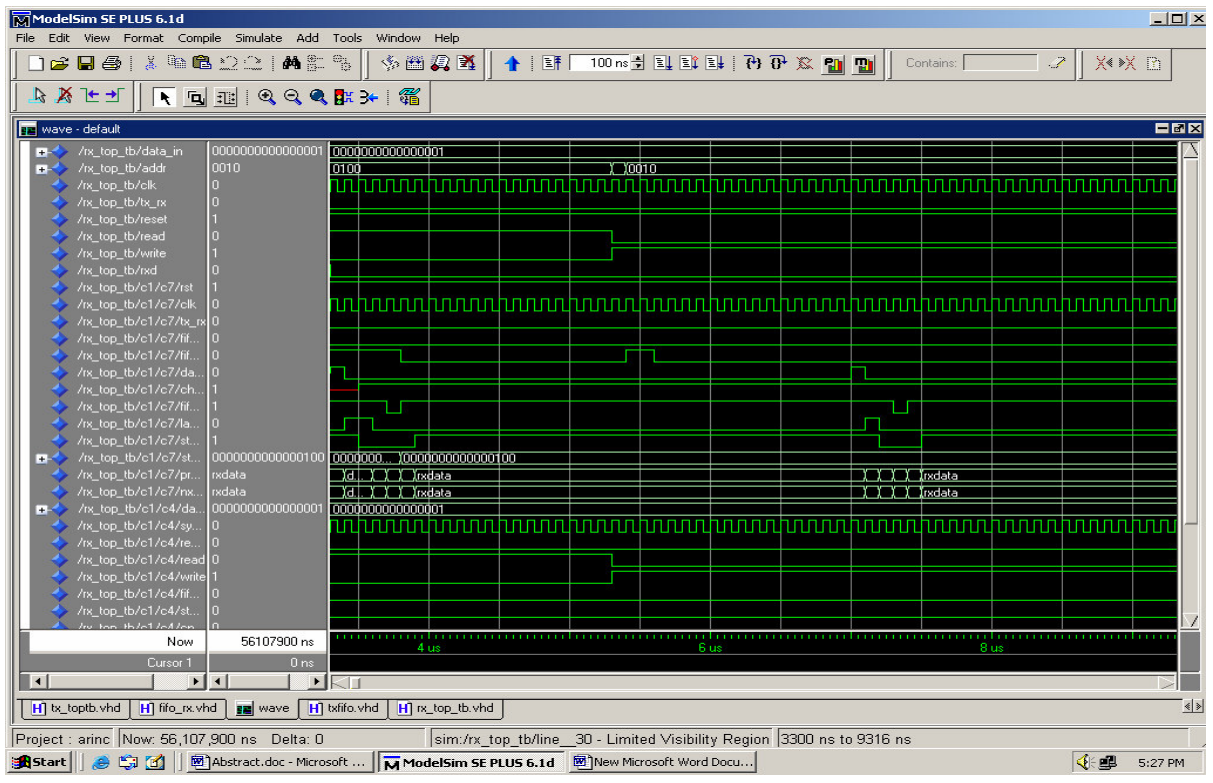

Fig.8a. Receiver Top Level Results 1

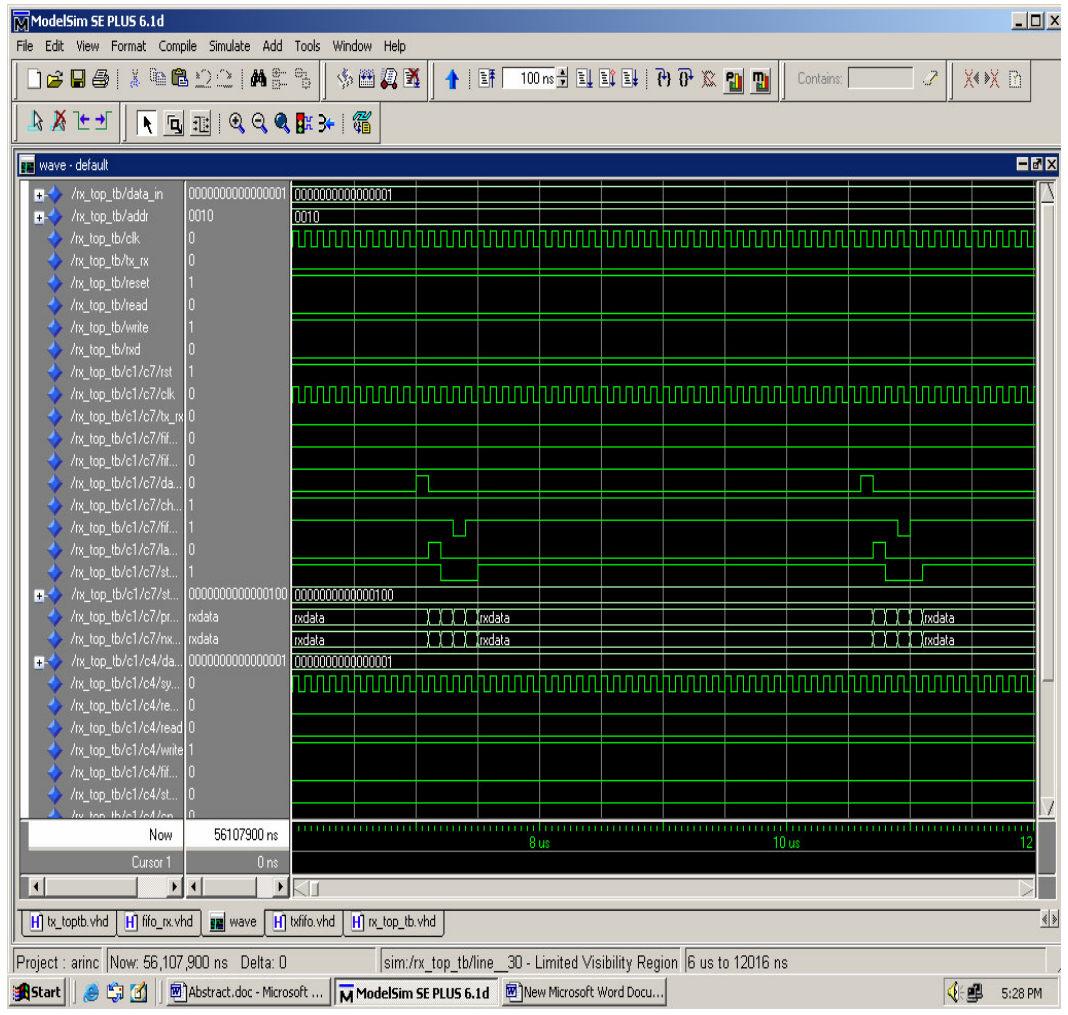

Fig.8b. Receiver Top Level Results 2

## 5.2 Flow summery,Floor planner,Chip view of ARINC core429

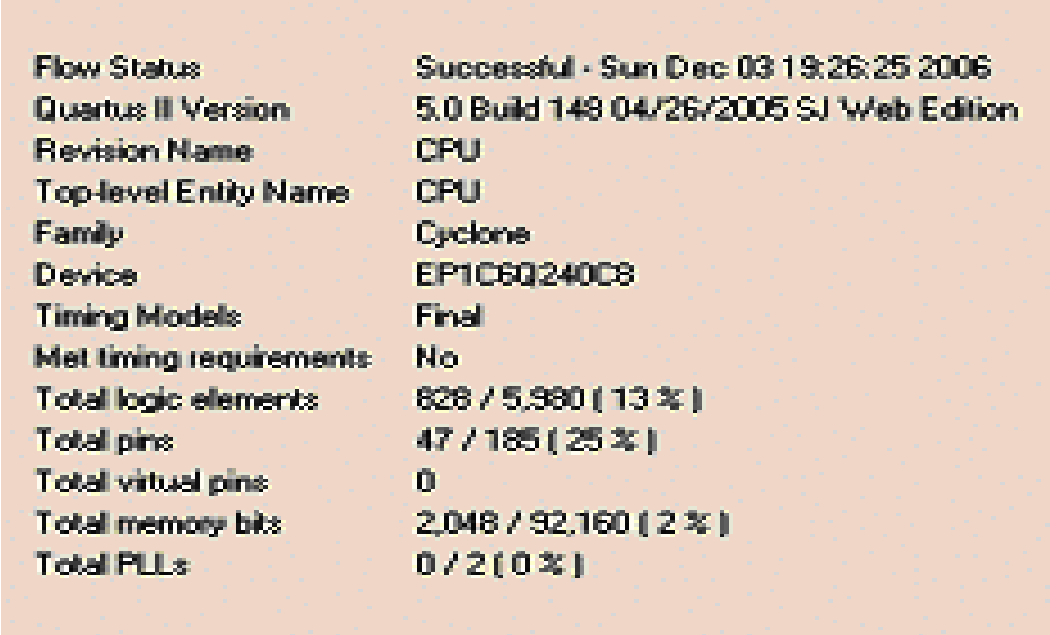

| | |
|---|---|
| Flow Status | Successful - Sun Dec 03 19:26:25 2006 |
| Quartus II Version | 5.0 Build 148 04/26/2005 SJ Web Edition |
| Revision Name | CPU |
| Top-level Entity Name | CPU |
| Family | Cyclone |
| Device | EP1C6Q240C8 |
| Timing Models | Final |
| Met timing requirements | No |
| Total logic elements | 828 / 5,980 ( 13 % ) |
| Total pins | 47 / 185 ( 25 % ) |
| Total virtual pins | 0 |
| Total memory bits | 2,048 / 92,160 ( 2 % ) |
| Total PLLs | 0 / 2 ( 0 % ) |

Fig 9 Flow summary of core 429

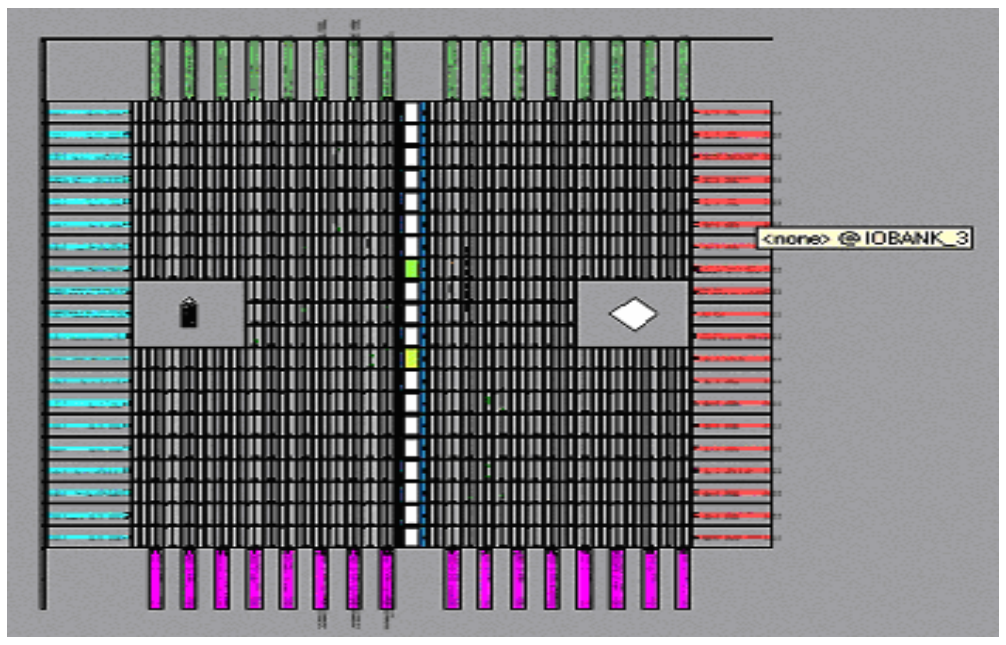

Fig. 10 Floor planner of core 429

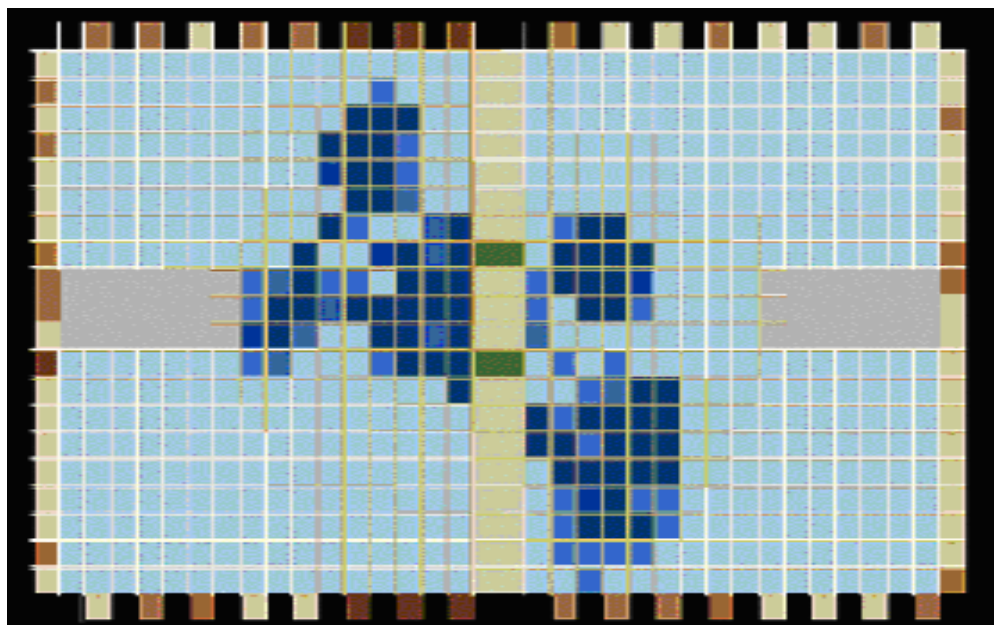

Fig. 11 chip view of core 429

## 5.3 RTL schematics of Transmitter, Receiver & core 429

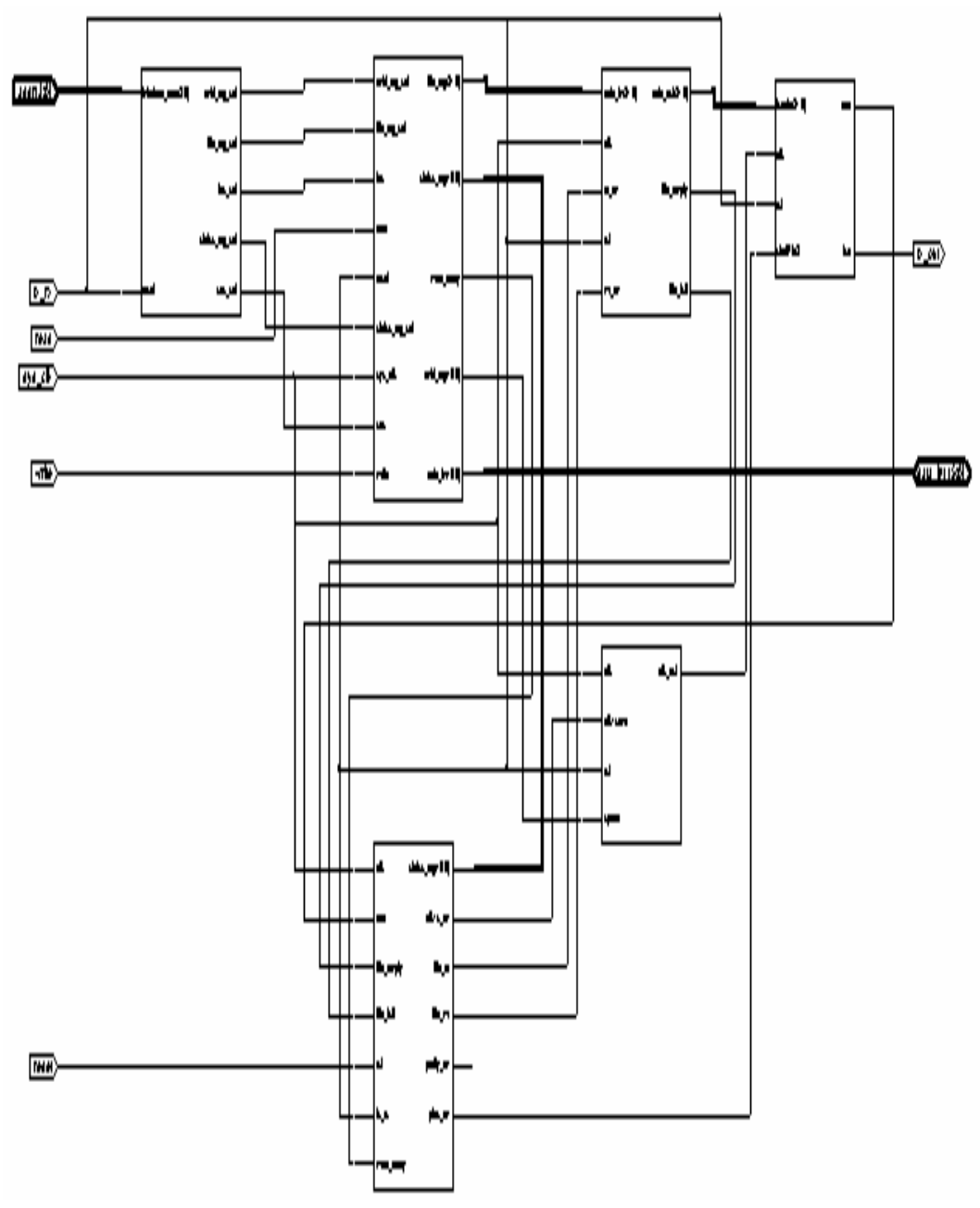

Fig.12 Transmitter top level RTL Viewer

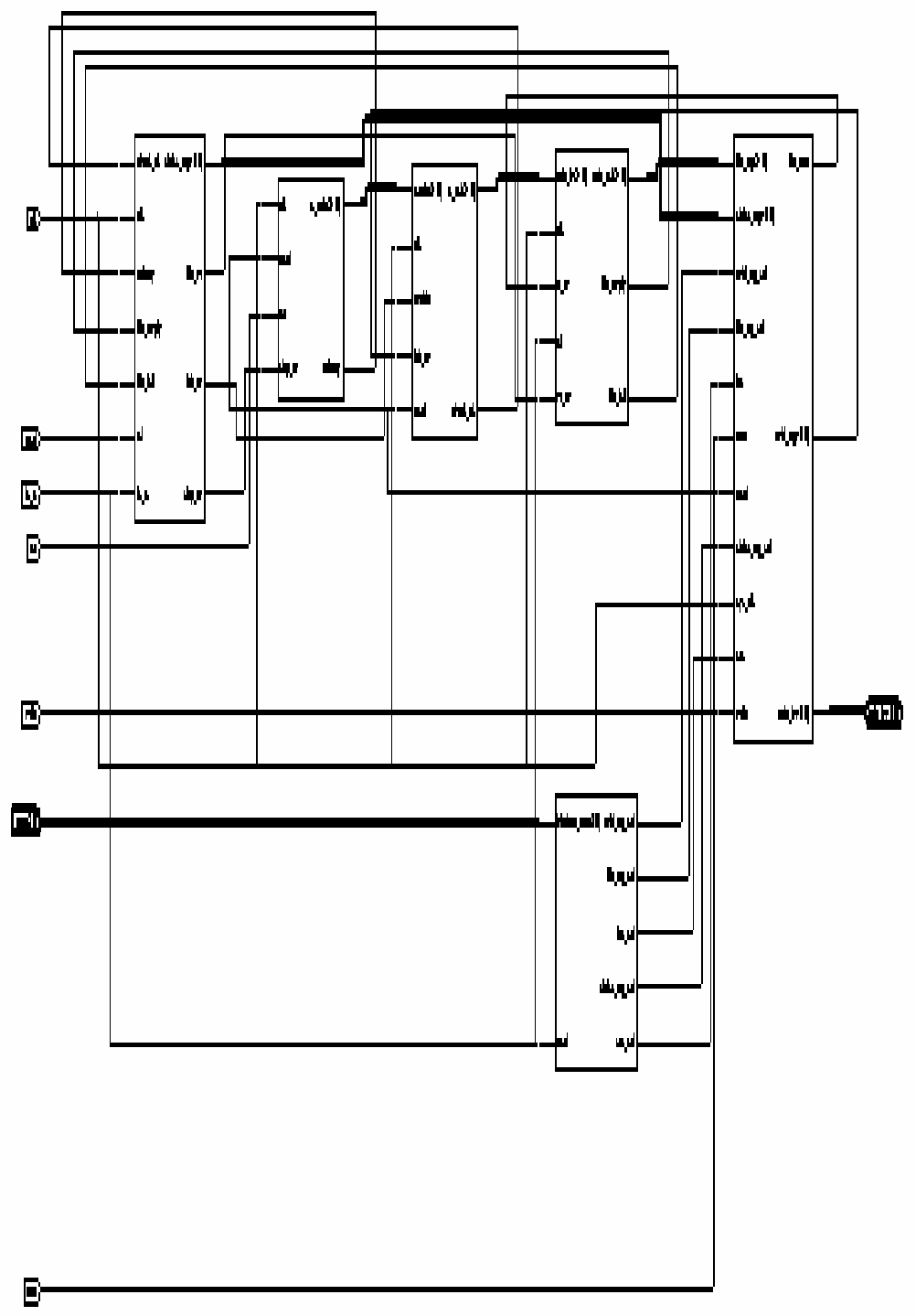

Fig. 13 Receiver top level RTL Viewer

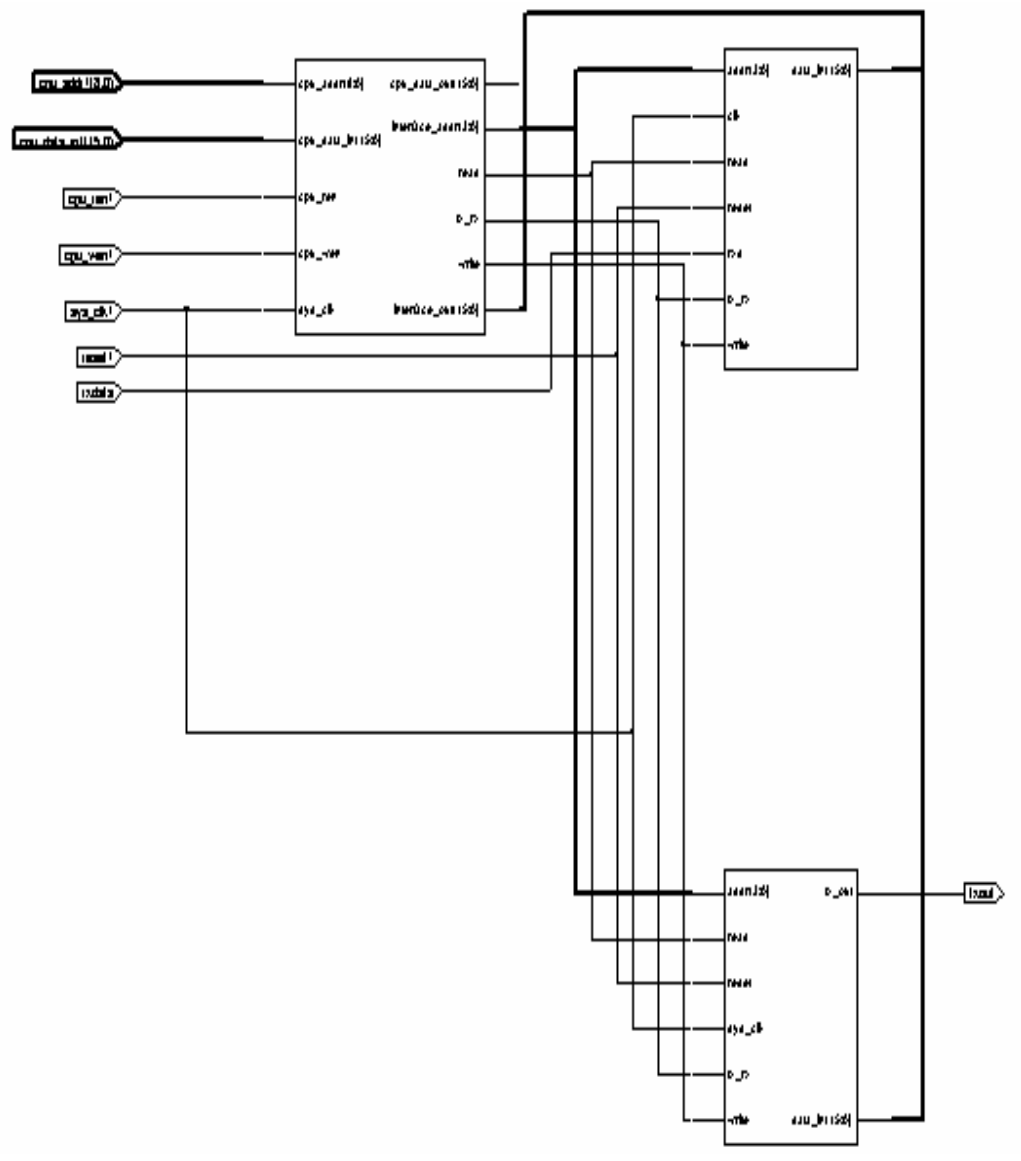

Fig.14 Core 429 RTL VIEWER

## 6.CONCLUSION

ARINC-429 is standard communication protocol used to link devices having this interface. In the course of executing project work the required hardware for data reception & transmission have been designed using VHDL and implemented on simulated FPGA. Provision has been given for 16 transmit/receiver channels but only one channel of Transmit /Receive is simulated. This can be duplicated for implementing channels in excess of one. The ARINC-429 data format are followed in implementing the logic.

Based on the simulation carried out, it is hereby conclude that the logic can be implemented in any standard FPGA device. Minor adjustments in Floor planning may be required with respective the FPGA manufacture by various vendors like Xilinx, Altera, Actel or any other sources.

An ASIC (Application Specific Integrated Circuit) may be made from the implemented logic. If the ASIC is made in numbers the cost of manufacturing can be reduced.

## 7.REFERENCES